\documentclass[pra,superscriptaddress,reprint,longbibliography]{revtex4-1}
\usepackage{graphicx}
\begin{document}

\title{Gendered Performance Differences in Introductory Physics: A Study from a Large Land-Grant University}

\date{\today}

\author{Matthew Dew}
\affiliation{Department of Physics and Astronomy, Texas A\&M University,
		College Station, TX~~77845}
\email[send correspondence to: ]{etanya@tamu.edu}
\author{Jonathan Perry}
\affiliation{Department of Physics, University of Texas,
		Austin, TX~~78712}
\author{Lewis Ford}
\affiliation{Department of Physics and Astronomy, Texas A\&M University,
		College Station, TX~~77845}
\author{William Bassichis}
\affiliation{Department of Physics and Astronomy, Texas A\&M University,
		College Station, TX~~77845}
		\author{Tatiana Erukhimova}
\affiliation{Department of Physics and Astronomy, Texas A\&M University,
		College Station, TX~~77845}

\begin{abstract}

Studies examining gender differences in introductory physics show a consensus when it comes to a gender gap on conceptual assessments; however, the story is not as clear when it comes to differences in gendered performance on exams. This study examined whether gendered differences exist on midterm and final exams in introductory physics courses and if such differences were correlated with a gender difference in final course grades. The population for this study included more than 10,000 students enrolled in algebra- and calculus-based introductory physics courses between spring 2007 and spring 2019.  We found a small but statistically significant difference, with a weak effect size, in final letter grades for only one out of four courses: algebra-based mechanics. By looking at midterm exam grades, statistically significant differences were noted for some exams in three out of four courses, with algebra-based electricity and magnetism being the exception. In all statistically significant cases, the effect size was small or weak, indicating that performance on exams and final letter grades was not strongly dependent on gender. As an added dimension examining gendered differences, we investigated if differences exist when accounting for instructor gender. Additionally, a questionnaire was administered in fall 2019 to more than 1,600 students in both introductory sequences to explore students' perceptions of performance, class contributions, and inclusion. We observed some differences between students' perception of their performance and contribution when grouped by gender, but no difference on perception of inclusion.

\end{abstract}

\maketitle

\section{Introduction}
Over the last half century, the number of US students majoring in STEM fields has more than doubled \cite{heron_mcneil_2016}. As this enrollment has increased, so has the attention paid to who is obtaining degrees across different disciplines, particularly when it comes to underrepresented groups. While some STEM disciplines, such as biology, have relative parity between males and females attaining degrees, other disciplines have a persisting gender gap \cite{AIPPaper}. The National Center for Science and Engineering Statistics found that in 2016, women earned 20.9\% of all engineering bachelor's degrees and 19.3\% of all physics degrees \cite{NSFPaper}.

Out of all STEM disciplines, physics is often considered to be the field that is the least welcoming for women to join \cite{1, 2}. Even for students not majoring in physics, STEM majors have to take physics as part of their academic program. For physics and engineering majors, introductory physics courses are among their early experiences in college. Such experiences can be crucial for student success within their majors \cite{3}.

The importance of such experiences is especially true for female students, as many of them leave physical science and engineering tracks during the first two years of college \cite{4}. Female students are likely to be under the pressure of gender stereotypes and societal biases \cite{5,6}, and they often find themselves underrepresented in their physics classes. Some authors argue that stereotype threat influences female student performance in introductory physics classes \cite{7, 8} and that use of an intervention based on value affirmation can help improve the situation \cite{9}. Perhaps related to stereotyping, the atmosphere in physics classrooms can influence female students’ physics self-efficacy, self-identity and motivation; all of which can have an impact on student success and retention \cite{10,11, 12, 13, Blue2019}. A number of studies have reported on the difference in physics self-efficacy between male and female students, including courses which use research supported instructional methods \cite{14, 15, 16, 17, Henderson}. As an example, Marshman et al. reported that female students had significantly lower self-efficacy than male students throughout a two-semester introductory physics course sequence \cite{18}. They go on to note that the physics self-efficacy of female students was negatively impacted by both traditional instruction courses and flipped classroom courses. 

A vast literature exists that explores the gendered differences in student performance on concept inventory tests in introductory physics courses. The majority of existing studies report a persistent gender gap with males performing significantly better than females on introductory mechanics concept inventory assessments \cite{FCI1, FCI2, FCI3, FCIFair1, FCIFair3, Fink2}, with some authors arguing that removing gender-biased context can reduce the gap \cite{FCIFair1, FCIFair2, FCIFair3}. The gender gap has been found in conceptual inventories of electromagnetism as well, although to a lesser extent and with more variation across studies \cite{BEMA1, FCI1, FCIFair3, Fink2, Fink3}. 

The results of prior studies on the gendered differences in student performance based on course grades and examinations are less consistent: while a number of studies indicate that male students outperform female students on the exams and course grades \cite{Fink1, 9, Matz}, other groups found no significant gendered difference in student performance \cite{FCI2, FCI3, FCI4, BEMA1, Lauer, Fink3, Fink2}. One study, comprised of 4,000 students across 7 semesters at the University of Colorado Boulder, reported a small but significant difference in course grades, correlated with differences in background factors for males and females \cite{Fink1}. Factors beyond the course, including prior knowledge, math background, and attitudes towards science, have been seen to correlate with gendered differences in performance \cite{Fink1, Weiman1}. One study of an electricity and magnetism course by Andersson and Johansson argues that the gendered difference in course grades disappears when controlled for the program in which a student is enrolled \cite{Andersson}. Tai and Sadler showed that females outperformed males in algebra-based courses while males outperformed females with the same background in calculus-based courses \cite{Tai}. Several studies performed on a large number of students taking the introductory physics classes report no significant gendered difference in student performance on course exams and course grades but a gender gap in concept inventories \cite{FCI2, FCI3, BEMA1}. Some studies of gendered differences in undergraduate physics have reported reduction or elimination through the use of carefully selected instructional strategies in introductory physics \cite{Mazur1, 17, Rodriguez}. However, other groups have found no effect of applying selected pedagogies or controlling the prior knowledge factors on gendered performance \cite{FCI1, Fink2, Karim2018, Cahill}. There have been a number of studies evaluating the impact of instructor gender on student outcomes in STEM and other fields and various metrics of their engagement \cite{bettinger2005faculty, Carrell2010, hoffmann2009professor, solanki2018looking}. Most studies reported small to no effect of having a same-sex instructor on course grades, except Carrell et al \cite{Carrell2010}. These studies pooled courses from different disciplines together and did not separately report the results on student performance in introductory physics courses.

This work focused on expanding the studies of academic gendered differences through large enrollment courses at Texas A\&M University. This is a large, land-grant institution, which yearly serves more than 20,000 undergraduate STEM majors across multiple colleges \cite{DarsTamu}. STEM majors at Texas A\&M complete their introductory physics sequence through either calculus-based courses (\emph{e.g.} physics, engineering, chemistry, math) or algebra-based courses (\emph{e.g.} life science majors, pre-meds, and environmental science). The engineering program comprises more than half of all STEM majors at Texas A\&M, so the calculus-based sequence enrollment is much larger than the algebra-based sequence. Students’ demographics, academic goals, and attitudes towards physics may differ significantly between calculus-based and algebra-based courses. Most students in the calculus-based courses were in their freshman year, i.e. right after high school. The algebra-based courses are typically taken by upper-level students in their sophomore to senior year, who do not have physics or physics-related disciplines as the focus of their studies and careers. Furthermore, there is a much larger proportion of female students in algebra-based courses. 

Our study aimed to examine gendered differences on both midterm and final exams as well as final letter grades for algebra- and calculus-based introductory physics courses at Texas A\&M. The objective of this study was to investigate whether gender differences on midterm and final exams existed and were correlated with a gender difference in final course grades. Prior literature reviewed above indicates that there is no consensus whether there is a gender difference in student final exam and course grades in the introductory physics classes. The question is still open and requires further investigation. This makes our study which includes a large data set spanning a long time period and 19 instructors particularly important as it reduces variability related to individual instructors and courses.  We analyzed all exam scores during the entire semester rather than the final grades only. We examined if statistically significant differences based on gender occurred on each exam for four introductory physics courses as the semesters were progressing, which has not previously been studied, at least not for such a large database and over a long period of time. To accomplish this task we created a database of course grades and scores from midterms and final exams for more than 10,000 students over a decade from two introductory physics course sequences: both calculus-based and algebra-based. As an added dimension to this study, we also looked at whether differences in student performance would be observed when separating by instructor gender. Beyond the database of grades mentioned above, we also took a snapshot of students' feelings via a short questionnaire to see how their perceptions aligned with historic performance.

\section{Methods}

From here forward, ``significant'' will be used as shorthand for ``statistically significant''. Statistical significance was taken to be at $p < 0.05$. In addition, tables will use ``Mech.'' or ``E\&M'' for mechanics or electricity and magnetism, respectively; ``Alg.'' or ``Calc.'' stand for algebra-based or calculus-based.

\subsection{Course Data}
To examine the gendered student performance within introductory courses, course level data were requested from faculty who taught one or more of these courses since 2007. Participating instructors provided students' first names, numerical scores for all midterm and final exams, and the final letter grade for the course. After collecting data from faculty, a database of approximately 13,000 students was obtained. This database contained information for students enrolled in the algebra-based sequence between 2011-2019 and the calculus-based sequence between 2007-2017. This study was structured in such a way that the only data collected were the course-level information provided by faculty. For this reason, connecting outcomes to non-academic factors was not possible with this study.

Since course-level data only included student names, gender was identified using an online tool, GenderizeIO \cite{GenderizeIO}. This application program interface uses census data to return a probability of gender based on the input of a first name and has been used in prior studies to attribute gender when these data were not available, e.g. Huang et al. 2020 \cite{huang2020historical}. Gender probability was considered identifiable for this study if it was 90\% or higher. This percentage was chosen as it allowed reasonable certainty of gender without drastically reducing the size of our data set. This cut eliminated about 17\% of our raw sample. The number of students identified as male or female from each of the four introductory courses examined in this study is shown in Table \ref{tab:GenderDist}.

\begin{table}[h]
\caption{\label{tab:GenderDist}Number of students and their gender distribution for each of the four introductory courses from the algebra- and calculus-based sequences.}
\begin{ruledtabular}
\begin{tabular}{llll}
& Total Number & Male & Female\\
Mech. Alg.  & 1,267              & 44.4\%      & 55.6\% \\
E\&M Alg.   & 999                & 44.6\%      & 55.4\%        \\
Mech. Calc. & 5,449              & 74.9\%      & 25.1\%        \\
E\&M Calc.  & 2,793              & 80.8\%      & 19.2\%
\end{tabular}
\end{ruledtabular}
\end{table}

\subsection{Comparing Grades}

\begin{figure}[h]
\centering
\includegraphics[width=8.6cm]{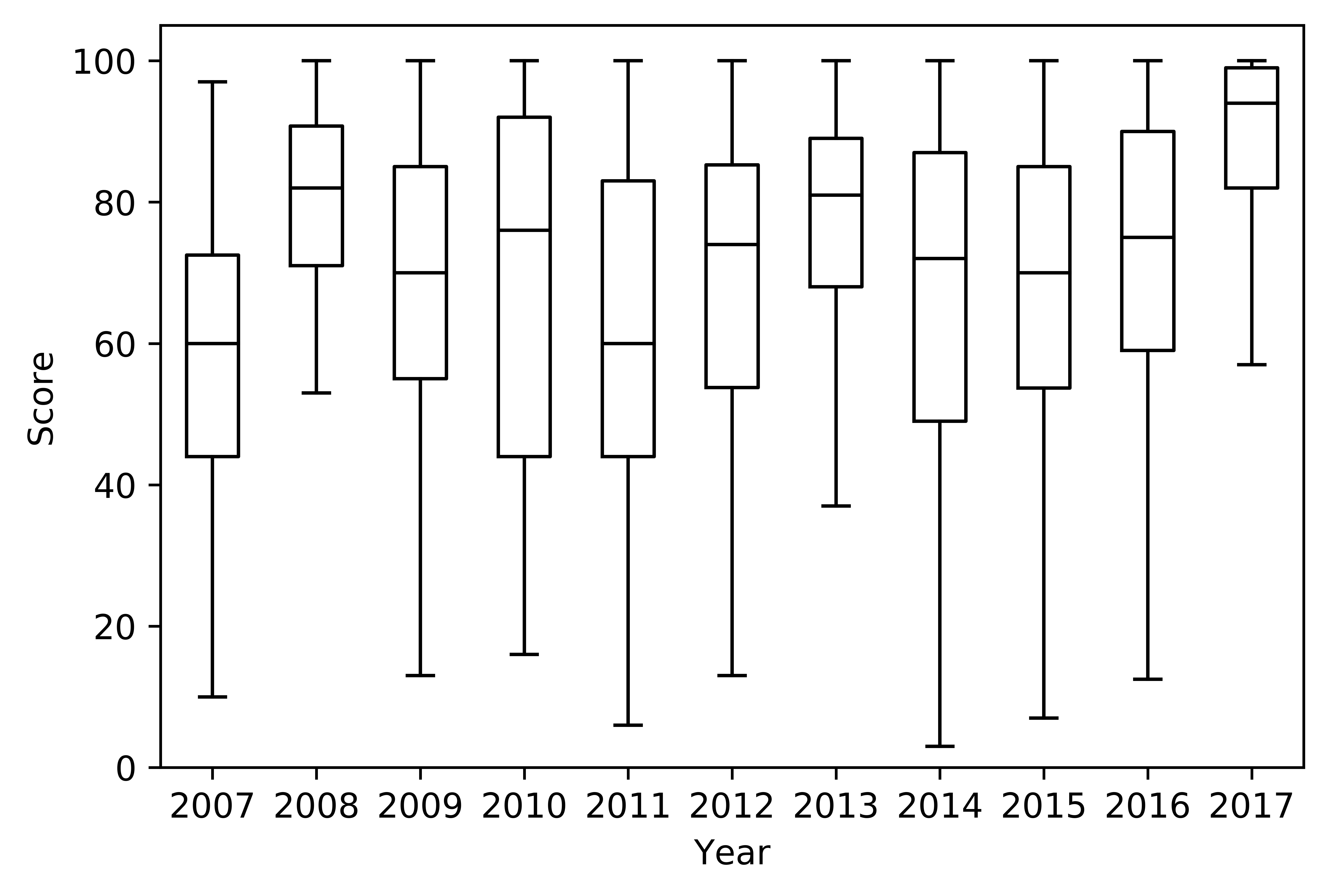}
\caption{\label{218boxplot}Raw score distribution of grades on the first calculus-based mechanics exam by year. For each box, the middle line represents the median, the box represents the two middle quartiles, and the error bars represent the highest and lowest quartiles.}
\end{figure}

Differences based on gender were examined by looking at students' final course grades and their scores on the midterm and final semester exams. Differences in populations were examined using t-tests between transformed data, as well as analysis of variance (ANOVA) applied to raw scores. These methods were used to examine the null hypothesis that there is no difference in performance between male and female students. Comparisons were made based on student gender, instructor, and year in the course. Some instructors gave multiple years of data, so these criteria allowed for individual lecture section distributions to be examined against each other. Though the exam distributions can be skewed for both raw scores and z-scores (see Figure \ref{218boxplot} for raw Exam 1 scores from calculus-based mechanics), violating the assumption of normality but not the assumption of homogeneity of variance according to Levene’s test, t-tests remained the most appropriate statistical analysis due to the large sample size from each course \cite{fagerland2012t}. Effect sizes were calculated using Cohen's $d$ with a Hedges correction  \cite{ResearchMethodsInEducation} with positive values indicating higher averages for male students and negative values indicating higher averages for female students. We consider $|d|<0.2$ to be weak, $0.2<|d|<0.5$ to be small, $0.5<|d|<0.8$ to be medium, and $|d|>0.8$ to be large effect sizes.

To look specifically at the relation between gender and exam performance, raw scores from individual lecture sections were mapped to new distributions using a z-score transformation. A z-score takes a raw numerical score ($x_i$), subtracts the average ($\bar{x}$), and scales by the standard deviation ($\sigma$), according to the relation:
\begin{equation}
    z = \frac{x_i - \bar{x}}{\sigma}. \label{eq:zscore}
\end{equation}
A positive z-score indicates how much higher a raw score was compared to the average in units of standard deviation. A negative z-score indicates the same but for a raw score below the average \cite{ResearchMethodsInEducation}. This transformation of scores was performed for a more even comparison of exam distributions across multiple years and instructors. For instance, Professor A teaching a course in year X might have a higher average and smaller deviation than when the same instructor teaches the same course in year Y. As an illustration of the z-score transformation, we can look at raw exam scores from calculus-based mechanics. In 2007, the mean score was 58 points and the standard deviation was 21 points. Students scoring a 58 were mapped to a z-score of 0, while students scoring a 37 were mapped to a z-score of -1. This was done for all students, using the individual lecture section averages and standard deviations.

Raw scores were used to examine individual lecture sections. When comparing across lecture sections, raw scores were transformed into z-scores so distributions may be more adequately and fairly compared to one another. Final course grades were treated on a 4 point scale (A-F) with no plus or minus grades per Texas A\&M's grading policy.

\subsection{Student Perceptions}
In fall 2019, a short anonymous questionnaire was administered to the algebra- and calculus-based courses to explore how students felt about their performance and inclusion. It must be noted that this was a questionnaire, rather than a validated survey instrument. The questionnaire was given during a semester that is outside the scope of the course level database described previously. We aimed to take a snapshot of students' perceptions to see how their responses aligned with historic data.

Students were asked to self-identify their race and gender. Response choices for students identifying as transgender or non-binary were available on the questionnaire. Only students identifying as male or female (99\%) were analyzed for this study. The questionnaire was composed of three questions where students responded using a 5-point Likert scale with responses that were negative (very and slightly), neutral, and positive (very and slightly): \newline
\begin{quote}
1. ``I felt included by my peers and instructors within this physics courses.'' \newline
2. ``I believe that I performed \underline{\hspace{0.4 in}} in this course.'' \newline
3. ``I felt that my contributions to discussions over physics material were valued during this course. This includes discussions both in class and outside of class but relating to completing assignments or preparing for exams.''
\end{quote}

The questionnaire was administered in the 12th and 13th weeks of a 15 week semester, which occurred mid-November in fall 2019. We chose these weeks as it was far enough into the semester that students would have formed an opinion of their inclusion but early enough that the questionnaires would not take away from finals preparation. The questionnaire was given during recitation, where students were in smaller groups and did not have their instructor in the room. The brevity of the questionnaire was dictated by the recitation format and an attempt to maximize the response rate. 

\section{Analysis and Results}
Differential performance for male and female students was examined for all exams and final course grades for four introductory physics courses. Results are separated by courses in the algebra-based sequence and the calculus-based sequence. The calculus-based sequence consists of three midterm exams administered throughout the semester with a comprehensive final (identified as Exam 4). The algebra-based sequence consists of four midterm exams administered throughout the semester with a comprehensive final (identified as Exam 5).

Questionnaire responses were converted into an ordinal 5-point scale, with higher numbers equating to more positive responses. That is,  better feelings of inclusion, greater performance, and stronger feelings of making valued contributions. 

\begin{table}[h]
\caption{\label{tab:FinalLetterGrades}Average final letter grades (and standard deviation) by gender for the algebra- and calculus-based introductory sequences, as well as the t-test and significance between these distributions.}
\begin{ruledtabular}
\begin{tabular}{lllll}
            & Male & Female & $t$ & $p$     \\ 
Mech. Alg.  & 2.687 (1.014)        & 2.839 (1.026)         & -2.634 & 0.009 \\ 
E\&M Alg.   & 2.935 (0.969)       & 2.875 (0.970)         & 0.967  & 0.334 \\ 
Mech. Calc. & 2.532 (1.172)       & 2.517 (1.108)         & 0.430   & 0.667 \\ 
E\&M Calc.  & 2.596 (1.119)       & 2.692 (1.080)         & -1.786 & 0.074 \\
\end{tabular}
\end{ruledtabular}
\end{table}

\subsection{Calculus-Based Mechanics and E$\&$M}

For calculus-based mechanics, data were provided by 14 instructors, for a total of 49 lecture sections. Eleven male instructors provided data from 34 lecture sections, and three female instructors provided data from 15 lecture sections. As seen in Table \ref{tab:FinalLetterGrades}, there was no significant difference observed when a t-test was applied to final letter grades based on student gender for the pooled data from all instructors and sections.

Gendered performance on course exams were compared using t-tests on transformed data from all instructors, Table \ref{tab:ExamCalcMech}. As a combined sample, male students scored higher to a significant degree on the first, third, and final exams of the course. The effect size of these differences is small ($d=0.240$) for the first exam, and weak for the third ($d=0.066$) and final exams ($d=0.105$). 

For calculus-based E\&M, data were provided by 10 instructors, for a total of 27 lecture sections. Six male instructors provided data from 10 lecture sections, and four female instructors provided data from 17 lecture sections. As seen in Table \ref{tab:FinalLetterGrades}, there was no significant difference observed when a t-test was applied to final letter grades based on student gender for the pooled data from all instructors and sections. 

As with the calculus-based mechanics course, exams were compared using t-tests on transformed data from all instructors, Table \ref{tab:ExamCalcEM}. No significant differences were observed between male and female students on any exam for this course.

As a validation of results found using t-tests on transformed data, three-way ANOVA was applied to raw scores for all exams from both calculus-based courses. Results were in agreement with t-tests applied to transformed data. Where significant differences were observed using t-tests, ANOVA showed gender to be a significant factor on its own or in combination with one or both of the other factors of professor and year. For the mechanics course, there were significant differences based on student gender for the first exam [F(1, 5401) = 51.21, p=9e-13], student gender and instructor for the third exam [F(13, 5401)=1.76, p=0.043], and student gender for the final exam [F(1, 5401)=5.77, p=0.016]. When examining individual lecture sections, significant differences due to gender were observed for less than 20\% of lecture sections. Combined with the results above, we note a persistent gender difference in calculus-based mechanics on exams only for pooled data, producing no significant difference in final course grades. No gendered differences were noted in calculus-based E\&M for either exams or final course grades. 

\begin{table}[h]
\caption{\label{tab:ExamCalcMech}Average exam z-scores (and standard deviation) for calculus-based mechanics, as well as the t-test and significance between these distributions.}
\begin{ruledtabular}
\begin{tabular}{lllll}
       & Male & Female & $t$ & $p$     \\
Exam 1 & 0.060 (0.987)      & -0.180 (1.016)       & 7.780   & \textless{}0.001 \\
Exam 2 & 0.007 (1.013)     & -0.020 (0.960)      & 0.858  & 0.391            \\ 
Exam 3 & 0.016 (1.012)     & -0.049 (0.963)     & 2.099  & 0.036            \\ 
Exam 4 & 0.026 (1.020)    & -0.077 (0.934)     & 3.305  & 0.001            \\
\end{tabular}
\end{ruledtabular}
\end{table}

\begin{table}[h]
\caption{\label{tab:ExamCalcEM}Average exam z-scores (and standard deviation) for calculus-based E\&M, as well as the t-test and significance between these distributions.}
\begin{ruledtabular}
\begin{tabular}{lllll}
       & Male & Female & $t$ & $p$     \\
Exam 1 & 0.013 (1.006)            & -0.053 (0.972)             & 1.360   & 0.174 \\
Exam 2 & 0.008 (1.002)            & -0.034 (0.991)             & 0.862  & 0.389 \\ 
Exam 3 & 0.005 (1.008)            & -0.020 (0.964)              & 0.502  & 0.616 \\ 
Exam 4 & 0.014 (1.001)            & -0.057 (0.992)             & 1.475  & 0.140  \\
\end{tabular}
\end{ruledtabular}
\end{table}

\subsection{Algebra-Based Mechanics and E$\&$M}

For algebra-based mechanics, data were provided by 4 instructors, covering a total of 13 lecture sections. Three male instructors provided data from 11 lecture sections, and one female instructor provided data from 2 lecture sections. Comparing the final letter grades based on student gender for pooled data from all instructors and sections showed a significant difference with a weak effect size ($d=-0.149$), Table \ref{tab:FinalLetterGrades}. This difference was only evident for pooled data. Only one lecture section exhibited a significant difference for letter grades. Removing this one section from pooled data, however, did not change the results.

Gendered performance on course exams for transformed data from all instructors for algebra-based mechanics is shown in Table \ref{tab:ExamAlgMech}. Significant differences were observed for the third and fourth exams with female students outperforming male students on average exam scores. The effect sizes for both the third ($d=-0.127$) and fourth ($d=-0.159$) exams were weak. 

For algebra-based E\&M, data were provided by 2 instructors, covering a total of 8 lecture sections. Two male instructors provided all of the data for this course. No significant difference in final letter grades based on student gender for pooled data from all instructors and sections was observed for this course, Table \ref{tab:FinalLetterGrades}. 

As with the algebra-based mechanics course, exams were compared using t-tests on transformed data from all instructors, Table \ref{tab:ExamAlgEM}. A significant difference in performance was observed only for the first exam with male students averaging higher than female students. The effect size for this difference was weak ($d=0.139$).  

As a validation of results found using t-tests on transformed data, three-way ANOVA was applied to raw scores for all exams from both algebra-based courses. Results were in agreement with t-tests applied to transformed data. Where significant differences were observed using t-tests, ANOVA showed gender to be a significant factor on its own or in combination with one or both of the other factors of professor and year. For the mechanics course, there were significant differences based on student gender for course grades [F(1, 1253)=5.65, p=0.018], on the third exam [F(1, 1253)=7.09, p=0.008], and the fourth exam [F(1, 1253)=10.87, p=0.001]. For the E\&M course, there was a significant difference based on student gender for the first exam [F(1, 993)=4.76, p=0.029].  When examining individual lecture section level data, significant differences due to gender were observed for less than 15\% of lecture sections. Combined with the results above, we note a gender difference in algebra-based mechanics on the third and fourth exams only for pooled data, producing a small but significant difference in final course grades. A gender difference is observed in algebra-based E\&M only for the first exam, producing no difference in final course grades.

\begin{table}[h]
\caption{\label{tab:ExamAlgMech}Average exam z-scores (and standard deviation) for algebra-based mechanics, as well as the t-test and significance between these distributions.}
\begin{ruledtabular}
\begin{tabular}{lllll}
       & Male & Female & $t$ & $p$     \\
Exam 1 & 0.012 (0.989)            & -0.010 (1.009)              & 0.396  & 0.692 \\
Exam 2 & -0.006 (0.993)           & 0.005 (1.005)              & -0.201 & 0.841 \\ 
Exam 3 & -0.071 (0.984)           & 0.056 (1.009)              & -2.248 & 0.025 \\ 
Exam 4 & -0.089 (1.021)           & 0.071 (0.977)              & -2.827 & 0.005 \\ 
Exam 5 & -0.029 (0.998)           & 0.023 (1.001)              & -0.923 & 0.356 \\
\end{tabular}
\end{ruledtabular}
\end{table}

\begin{table}[h]
\caption{\label{tab:ExamAlgEM}Average exam z-scores (and standard deviation) for algebra-based E\&M, as well as the t-test and significance between these distributions.}
\begin{ruledtabular}
\begin{tabular}{lllll}
       & Male & Female & $t$ & $p$     \\ 
Exam 1 & 0.077 (0.986)            & -0.062 (1.007)             & 2.182  & 0.029 \\ 
Exam 2 & 0.053 (1.012)            & -0.042 (0.988)             & 1.492  & 0.136 \\ 
Exam 3 & -0.008 (0.946)           & 0.007 (1.042)              & -0.235 & 0.814 \\ 
Exam 4 & 0.023 (1.031)            & -0.019 (0.974)             & 0.653  & 0.514 \\
Exam 5 & 0.022 (0.992)            & -0.018 (1.006)             & 0.624  & 0.533 \\
\end{tabular}
\end{ruledtabular}
\end{table}

\subsection{Instructor Gender}
The impact of instructor gender on differences in student performance by gender was also examined. This analysis could not be applied to the algebra-based course sequence as there were data available from only one female instructor for the mechanics course and no data available from female instructors for the E\&M course. Data were separated into two groups by instructor gender, and comparisons were made based on student gender.  Differences in average z-scores for each course exam are shown in Table \ref{tab:InstructorZScores}, while comparisons between letter grades are shown in Table \ref{tab:InstructorLetterGrades}. 

For calculus-based mechanics, data were obtained from eleven male instructors for 34 lecture sections (N=4,227) and from three female instructors for 15 lecture sections (N=1,222). Significant differences in student performance based on gender were observed for instructors of both genders on the first midterm exam, with a small effect size for male ($d=0.229$) and female ($d=0.243$) instructors. Other significant differences were noted for the third and fourth exams for male instructors only. These latter two differences have weak effect sizes ($d=0.085$ and $d=0.117$, respectively). No significant difference was noted in final letter grades when grouping students by instructor gender. 

For calculus-based E\&M, data were obtained from six male instructors for 10 lecture sections (N=917) and from four female instructors for 17 lecture sections (N=1,876). A significant difference in student performance based on gender was observed for the fourth exam of the semester for male instructors. This difference has a small effect size ($d=0.304$). A significant difference was also observed in letter grades for female instructors, with a weak effect size ($d=-0.154$). 

We also analyzed the raw score data using ANOVA. Agreement on significance using t-tests and ANOVA was found for comparisons based on instructor gender except for two instances. These instances were calculus-based mechanics on the first exam for female instructors [F(1, 5445)=3.39, p=0.065] and calculus-based E\&M on the final exam for male instructors [F(1, 2789)=3.08, p=0.079]. For the calculus-based mechanics course, there were significant differences based on student gender for male instructors for the first exam [F(1, 5445)=21.2, p=4e-6], third exam [F(1, 5445)=7.08, p=0.008], and the final exam [F(1, 5445)=7.24, p=0.007]. Further examination of the impact of instructor gender was performed using Tukey HSD, which found significance in agreement with t-tests \cite{ResearchMethodsInEducation}.

\begin{table*}[h]
\caption{\label{tab:InstructorZScores}Average z-score differences between male and female students, $\Delta$, separated by instructor gender, as well as the t-tests and significances between these distributions.}
\begin{ruledtabular}
\begin{tabular}{lllllll}
  & \multicolumn{3}{l}{Male Instructors} &  \multicolumn{3}{l}{Female Instructors}                  \\ \hline
  & $\Delta$ & $t$ & $p$ & $\Delta $         & $t$ & $p$                \\ \hline
Mech. Calc.  &          &        & &                  &        &                  \\
Exam 1      & 0.242    & 6.921  & \textless{}0.001 & 0.233            & 3.554  & \textless{}0.001 \\
Exam 2      & 0.019    & 0.532  & 0.595 & 0.054            & 0.826  & 0.409            \\
Exam 3      & 0.084    & 2.387  & 0.017 & -0.001 & -0.010  & 0.992            \\
Exam 4      & 0.115    & 3.260   & 0.001 & 0.060             & 0.912  & 0.362           \\ \hline
E\&M Calc. &          &        &  &                  &        &                  \\
Exam 1      & 0.128    & 1.404  & 0.161 & 0.041            & 0.731  & 0.465            \\
Exam 2      & 0.054    & 0.595  & 0.552 & 0.037            & 0.647  & 0.518            \\
Exam 3      & 0.034    & 0.370   & 0.711 & 0.021             & 0.362  & 0.717            \\
Exam 4      & 0.307    & 3.384  & 0.001 & -0.020            & -0.352 & 0.725            
\end{tabular}
\end{ruledtabular}
\end{table*}

\begin{table*}[h]
\caption{\label{tab:InstructorLetterGrades}Average final letter grade difference between male and female students, $\Delta$, separated by instructor gender, as well as the t-tests and significances between these distributions.}
\begin{ruledtabular}
\begin{tabular}{lllllll}
 & \multicolumn{3}{l}{Male Instructors} & \multicolumn{3}{l}{Female Instructors} \\ \hline
Class & $\Delta$ & $t$ & $p$ & $\Delta$ & $t$ & $p$     \\
Mech. Calc.  & 0.012    & 0.282  & 0.778 & 0.031     & 0.449  & 0.654 \\
E\&M Calc. & 0.141    & 1.351  & 0.177 & -0.167   & -2.693 & 0.007
\end{tabular}
\end{ruledtabular}
\end{table*}

\subsection{Students’ Perception Questionnaire}
A short questionnaire consisting of three questions, that were analyzed independently, as well as demographic information was distributed to the students taking introductory physics classes in the fall of 2019, as described in Section II C. We received over 1,600 completed questionnaires with a response rate of 63\%. 

On the question about students’ perception of their performance,

\begin{quote}
2. ``I believe that I performed \underline{\hspace{0.4 in}} in this course.'' 
\end{quote}
they were given five answer choices:
\begin{quote}
A. Well Below Average \newline 
B. Below Average \newline 
C. Average \newline 
D. Above Average \newline 
E. Well Above Average
\end{quote}
We converted these answers into a 5-point scale with ``Well Below Average'' corresponding to a 1 and ``Well Above Average'' to a 5. 

Using the ordinal scale described above, t-tests were applied to examine the null hypothesis for student perceptions. We found that female students rated their performance perception equally to male students only in algebra-based mechanics (See Table \ref{tab:PerformancePerception}). When compared to historic data, this is the one course examined where we found a statistically significant difference in final letter grades with female students outperforming male students.

For the other three courses, male students rated their performance as significantly higher with effect sizes ranging from weak to medium. In the calculus-based mechanics course, male students rated their performance a third of a point higher than their female classmates as shown in Table \ref{tab:PerformancePerception} ($p < 0.001$); the effect size was small ($d=0.389$). In the algebra- and calculus-based E\&M courses, male students rated their performance higher than their female classmates ($p < 0.05$) with a medium effect size ($d=0.504$) for algebra-based and a weak effect size for calculus-based ($d=0.168$). For both E\&M courses, male students performed the same as female students at this point in the course according to historic data. 

\begin{table}[h]
\caption{\label{tab:PerformancePerception}Average student performance perception (and standard deviation) for each course, as well as the t-test and significance between these distributions. }
\begin{ruledtabular}
\begin{tabular}{lllll}
&Male & Female & $t$ & $p$ \\
Mech. Alg.  & 3.384 (0.974) & 3.384 (0.873) & -0.003 & 0.998 \\
E\&M Alg.   & 3.342 (0.867) & 2.886 (0.919) & 2.154  & 0.035 \\
Mech. Calc. & 3.395 (0.883) & 3.062 (0.824) & 3.699  & \textless{}0.001 \\
E\&M Calc.  & 3.545 (0.891) & 3.391 (0.938) & 2.000  & 0.046
\end{tabular}
\end{ruledtabular}
\end{table}

Next, we looked at how students rated their feelings of inclusion. More specifically, we asked the following,
\begin{quote}
1. ``I felt included by my peers and instructors within this physics course.'' 
\end{quote}

Finally, we asked students whether they felt that their contributions were valued. 
\begin{quote}
3. ``I felt that my contributions to discussions over physics material were valued during this course. This includes discussions both in class and outside of class but relating to completing assignments or preparing for exams.'' 
\end{quote}
For both of these questions, the students were given the following answer choices:
\begin{quote}
A. Strongly Disagree \newline 
B. Disagree \newline 
C. Neither Agree nor Disagree \newline 
D. Agree \newline 
E. Strongly Agree
\end{quote}

Similar to the question on performance perception, we converted these answers into a 5-point scale with ``Strongly Disagree'' corresponding to a 1 and ``Strongly Agree'' to a 5. 
\begin{table}[h]
\caption{\label{tab:FeelingsInclusion}Average student feelings of inclusion (and standard deviation) for each course, as well as the t-test and significance between these distributions.}
\begin{ruledtabular}
\begin{tabular}{lllll}
&Male & Female & $t$ & $p$ \\
Mech. Alg.  & 3.767 (1.340) & 3.813 (1.311) & -0.318 & 0.751 \\
E\&M Alg.   & 4.026 (1.158) & 4.229 (0.865) & -0.828 & 0.410  \\
Mech. Calc. & 3.994 (1.155) & 4.031 (1.037) & -0.316 & 0.752 \\
E\&M Calc.  & 3.964 (1.149) & 4.005 (1.090) & -0.431 & 0.667 
\end{tabular}
\end{ruledtabular}
\end{table}

We found that despite a difference in performance perception, there was no statistically significant difference in feelings of inclusion for any of the courses (See Table \ref{tab:FeelingsInclusion}). That is, despite female students believing they were underperforming in three courses, they still believed they were being included equally. 

\begin{table}[h]
\caption{\label{tab:Contribution}Average student contribution valuation (and standard deviation) for each course, as well as the t-test and significance between these distributions.}
\begin{ruledtabular}
\begin{tabular}{llllll}
&Male & Female & $t$ & $p$ \\
Mech. Alg.  & 3.555 (0.951) & 3.793 (0.793) & -2.52 & 0.012 \\
E\&M Alg.   & 3.474 (1.019) & 3.343 (1.120) & 0.515 & 0.608 \\
Mech. Calc. & 3.727 (0.906) & 3.700 (0.891) & 0.289 & 0.773 \\
E\&M Calc.  & 3.803 (0.891) & 3.769 (0.907) & 0.453 & 0.650 
\end{tabular}
\end{ruledtabular}
\end{table}

When analyzing whether students felt that their contributions were valued, we found a statistically significant difference in algebra-based mechanics (See Table \ref{tab:Contribution}). This had a small effect size ($d=-0.271$). In this course, female students rated valuations of their contributions about a fourth of a point higher on average. This corresponds to the only course where female students historically performed better on final letter grades. We found no significant differences in the other courses, indicating male and female students had similar feelings about how their contributions were valued.

To substantiate our methods of analyzing questions separately, we calculated the Spearman rank-order correlation coefficient between each questionnaire item to determine the strength of correlation. We found that each question's correlations ranged from a small effect to a medium effect ($0.2<\rho<0.45$). This includes when we sorted by gender, course, and simultaneously gender and course.

\section{Discussion}
We compared student outcomes for final course grades and exams based on gender for over 10,000 Texas A\&M students over more than a decade. We examined this data to determine whether such differences were persistent throughout each course. These data were collected from instructors who taught courses from algebra-based or calculus-based introductory physics sequences. Prior studies of gendered students’ performance based on exam grades show mixed results, with some authors reporting that male students outperform their female counterparts \cite{Fink1, 9, Matz}, whereas other authors did not find a statistically significant difference between the genders \cite{FCI2, FCI3, FCI4, BEMA1, Lauer, Fink3, Fink2}. To describe our results, we use ``significant'' as a shorthand for ``statistically significant'', defined as $p < 0.05$. 

Of the four introductory courses comprising the algebra-based and calculus-based sequences, only the algebra-based mechanics course exhibited a significant difference in final letter grades. This difference is small, with female students outperforming male students by 0.15 GPA points and has a weak effect size. This is in agreement with a prior study by Tai and Sadler who also reported female students performing better than male students in algebra-based mechanics \cite{Tai}. In algebra-based mechanics, significant performance differences based on gender were found on two out of five exams throughout the course. These were midterm exams in the second half of the course, and the differences had weak effect sizes.

In calculus-based mechanics, there was no significant difference in final letter grades. There was, however, a significant difference on three out of four course exams, with a small effect size for the first exam, and weak effect sizes for the third and fourth exams. Previous studies have found inconsistent results from calculus-based mechanics courses. Some studies conducted at other public, state universities have found no significant gap in final letter grades \cite{FCI2, Fink2, Lauer}. Kost et al. found a small but statistically significant difference in overall course grades \cite{Fink1}. Also, Tai and Sadler cited above for the algebra-based course, reported males outperforming females in calculus-based mechanics \cite{Tai}.

In the second semester E\&M course for both sequences, no significant difference in final letter grades was found. The only significant difference observed was for the first exam in the algebra-based course which had a weak effect size. The results are similar to Kost-Smith et al. \cite{Fink3} who also found no statistically significant difference in course grades in a calculus-based electricity and magnetism course.

When examining the relationship between instructor gender and student gendered performance, fewer significant differences were observed for female instructors in comparison with male instructors. Where significant differences were seen, the effect sizes were small or weak, similar to the effect sizes when data from all instructors were pooled. These results are in agreement with prior studies \cite{hoffmann2009professor, solanki2018looking} across multiple disciplines examining the impact of instructor gender.

In addition to analyzing our data set of historic exam scores between fall 2007 and spring 2019, we took a snapshot of students’ perception of their performance as well as their feelings of inclusion and contribution during fall 2019. We administered a questionnaire to all students who took introductory physics classes in this semester. We received over 1600 completed questionnaires with a response rate of 63\%. It should be noted that due to the timing of the questionnaire responses were not linked to course performance.

Responses indicated that female students had lower perception of their performance than their male classmates with effect sizes ranging from weak to moderate. The one exception is algebra-based mechanics where female students rated their perception of performance equal to male students. Our question about the performance perception didn't directly explore students’ physics self-efficacy. However, we believe that our results complement previous studies reporting that female students display lower self-efficacy than male students in introductory physics classes, even when controlled for their performance level \cite{18, 10, 14, 15, 16, 17}.

We found no significant gendered difference in students' perceptions of inclusion across all courses. For three courses there were no significant differences in students' perceptions about the value of their contributions. In algebra-based mechanics, female students reported their contributions as valued higher than male students with a small effect size. Gender neutral perception of inclusion and perception about the value of students’ contributions is positive news taking into account that the questionnaire was distributed towards the end of the semester when students had enough time to form an opinion. In algebra-based courses, the enrollment of students tends to have more female students than male students, and students are typically upper-level undergraduates. As a result, female students in these classes may experience less stereotype threat \cite{8}, which could partially explain why they report equal or better perception of inclusion and valuation of their contribution as compared to their male counterparts. In algebra-based mechanics, the instruction during the recitations was provided by upper-level undergraduate students who happened to be about $70\%$ females in fall 2019 when the questionnaire was distributed. This could contribute to female students rating the value of their contributions higher than their male classmates \cite{12}. It is also worth noting that female students rated their performance as equal to male students and the value of their contributions higher than male students in the algebra-based mechanics class, where female students historically outperformed male students based on our analyses of more than a decade of exam data.

\section{Limitations}
While this study has the advantage of a large data set of exam scores and final course grades collected over a decade and provides new knowledge on the gendered performance in introductory physics classes, it has some limitations. The most significant of which is only course-level data collected from faculty were used in this study. Therefore, we did not analyze the impact of non-academic factors that have been seen to potentially account for 20\% to 70\% of the gender difference \cite{Fink1, Weiman1}. Additionally, since instructors did not store student gender in their data, we determined it using first names similar to previous studies \cite{huang2020historical}.

One should keep in mind that the instructors who taught these classes used different instruction methods that were evolving during the period of data collected. The student demographics, background, and preparation levels were evolving during the years of study \cite{DarsTamu}. Since this is a long-term historic study of course-level data, we were not able to control for factors such as student background in math and physics. It should be noted that the overall number of instructors included in this study was small, 19, and a larger sample of instructors would help to determine the consistency of the impact of instructor gender on student gendered performance. Also, we had a smaller subset of female instructors than male instructors.

The questionnaire distributed in the fall 2019 semester was meant to provide a snapshot of students' feelings and was not a validated survey instrument. The questionnaire was administered during a semester outside the years where course-level data were collected. 

\section{Conclusion}
This study explored gender differences in student performance on exams and final letter grades for algebra- and calculus-based introductory physics sequences. The performance indicators for a large pool of over 10,000 students spanning a period from spring 2007 to spring 2019 have been analyzed. Data on midterm exams, final exams, and final course grades were collected from instructors teaching these courses during that period of time. Our goal was to investigate whether there was a measurable gender difference on midterm and final exam scores and if they were correlated with gender differences in final course grades. By utilizing a large data set, differences due to individual instructors or other transient factors were averaged out. Where differences in final letter grades were found for a course, there were no persistent differences observed across that course's exams. Where persistent differences were observed on exams within a course, there were no differences for final letter grades in that course. In algebra-based mechanics, female students outperformed male students by a small but statistically significant margin. In all statistically significant cases, the effect size was small or weak, indicating that performance on exams and final letter grades was, at most, weakly dependent on gender. While not the main focus of our study, we looked at the relationship between instructor gender and student performance and observed fewer statistically significant differences for students with female instructors.

Our paper provides new data on the gender difference on exams and course grades within introductory physics through a sizable data set collected from more than ten years of courses at a large public university. Prior to this study, we considered it an open question whether significant gendered differences could appear on particular exams but not be large enough to affect final course grades. Our findings provide new information that performance on exams and final letter grades are weakly dependent on student gender. These results may help with fighting the gender stereotypes that negatively impact so many female students \cite{8}.

A questionnaire was distributed to students taking both calculus-based and algebra-based sequences during the fall 2019 semester. The goal was to take a snapshot of current students' feelings to see how their perceptions aligned with historic performance. We collected students' feedback on their perception of their performance, feelings of inclusion, and the value of their contributions. The analyses of student responses revealed no difference in the feeling of inclusion in any of these courses. For one course, algebra-based mechanics, female students rated their contributions as valued more compared to male students. For the same course, female students reported their performance perception to be as high as their male counterparts. For the other three courses, male students reported higher perceptions of performance than female students. 

There are several future studies that could stem from this one. In the next iteration of this work, it would be beneficial to connect course-level data with university records of students’ prior preparation and knowledge. While gender was not a strong factor leading to differences in student performance across combined samples, a minority of sections did exhibit differences. A subsequent study could examine individual course sections which exhibit statistically significant differences based on student gender for the factors which might contribute to these differences. Additionally, an enhanced survey on student perceptions could be linked with course performance to allow for regression analyses between inclusion and contribution with success on exams. Since calculus-based mechanics is usually taken the earliest of these four courses, it would be useful to perform a study like this one on calculus 1 and introductory chemistry. This would help us better understand if gendered performance differences among physical science and engineering majors change over time.

\begin{acknowledgments}
This study was supported by the Texas A\&M University College of Science. We would like to thank the Texas A\&M University Department of Physics and Astronomy faculty who provided us with data for this study. 
\end{acknowledgments}

\bibliography{bibliography}

\end{document}